\documentclass[11pt]{article}
\usepackage{amsmath,amssymb}
\usepackage{graphicx}
\usepackage{multirow}
\usepackage{subeqnarray}
\usepackage{cases}
\usepackage{titlesec}
\usepackage[titletoc]{appendix}
\usepackage{algorithm}
\usepackage{algorithmic}
\usepackage{multirow}
\usepackage{xcolor}
\usepackage{bm}
\usepackage{booktabs}

\newtheorem{definition}{Definition}
\newtheorem{theoremx}{Theorem}
\newtheorem{lemmax}{Lemma}

\newtheorem{assumption}{Assumption}

\begin{document}
\title{\textbf{A hybrid approach for cooperative output regulation with sampled compensator and its application in micro-grid
\thanks{ This work was supported in part by the National Natural
Science Foundation of China under Grant 61272069 and 61272114.}} \vspace{1cm%
}}

\author{\textrm{\textbf{ Chao Yang$^1$, Zhi-Hong Guan$^{1,}$\thanks{Corresponding author. E-mail address:
zhguan@mail.hust.edu.cn(Z.-H.Guan).}, Ming Chi$^1$, Gui-Lin Zheng$^2$}} \\
\\
{\small \textsl{$^1$College of Automation, Huazhong University of Science and Technology,
Wuhan, 430074, P. R. China}}\\
{\small \textsl{$^2$Department of Automation, School of Power and mechanical engineering}}\\
{\small \textsl{Wuhan University,
Wuhan,
430072, P. R. China}}
}
\date{}
\maketitle

\begin{abstract}
    This work investigates the cooperative output regulation problem of linear multi-agent systems with hybrid sampled data control. Due to the limited data sensing and communication, in many practical situations, only sampled data are available for the cooperation of multi-agent systems. To overcome this problem, a distributed hybrid controller is presented for the cooperative output regulation, and cooperative output regulation is achieved by well designed state feedback law. Then it proposed a method for the designing of sampled data controller to solve the cooperative output regulation problem with continuous linear systems and discrete-time communication data. Finally, numerical simulation example for cooperative tracking and a simulation example for optimal control of micro-grids are proposed to illustrate the result of the sampled data control law.
    \vspace{1em}
\end{abstract}

\section{ Introduction }
Since the beginning of this century, cooperative control problems of multi-agent systems have received considerable attention, such as consensus, formation, flocking and rendezvous problems etc. \cite{Saber2004,Saber2006,Egerstedt2009,Liu2012}.
Since coordinating autonomous agents can achieve some challenging tasks which are impossible for single agent, cooperative control of multi-agent systems has been applied in many areas, for instants, formation of spacecraft, scheduling of traffic systems and optimal dispatch of distributed generation \cite{Ren2004,Krasemann2012,zhang2012}.
Besides consensus, formation and flocking, various cooperative control techniques has been developed, such as optimal coordination \cite{F2014GameA,Qian2014},  cooperative learning \cite{La,Papa2013} and cooperative output regulation problem \cite{huangjie2012}.

Output regulation is a fundamental problem in control systems, there are some classic works studied this problem \cite{Francis1977,Huang2004,Byrnes2012}.
It aims to meet requirement of output tracking and disturbance rejection of controlled plants. The reference signal for tracking or the disturbance for rejection are generated by some exosystems.

Recently, output regulation problem for multi-agent systems receives much attention due to their advantage in dealing with cooperation problem of heterogeneous plants. Xiang et al applied output regulation on the synchronization control of identical subsystems \cite{Xiang2009}, in this work, output of each subsystems was regulated and tracked the signal converged to synchronous manifold. Then in \cite{huang2011}, a more relaxable condition for these systems was carried out. In \cite{huangjie2012}, a progressive method was developed to solve the output regulation of heterogeneous multi-agent systems by Su and Huang. Wieland, et.al. proposed a control method to solve the problem of leaderless consensus for heterogeneous plants. They handled the problem by setting up a virtual leader and what worth to mention they gave a necessary and sufficient condition for the linear output synchronization of heterogeneous multi-agent systems. There were lots of worthy works studied this problem in different perspectives, some considered the uncertainty of subsystems \cite{su2013,Liang2014}, others considered the switching of network topology \cite{su2012}. Moreover, there were numerous works studied output regulation for identical or heterogeneous nonlinear systems \cite{Dong2014,Xu2014,Isidori2014}.

This paper aims to study  the cooperative output regulation with sampled data information from neighboring agents and exosystem.
Generally, cooperation of autonomous agents are physical plants coordinating with each other via cyber networks (communication or sensing networks). Thus most multi-agent systems are typical cyber-physical systems (CPSs) \cite{Sztipanovits}. A general modeling framework for CPSs is hybrid system \cite{Derler}. In coordination, the task and role assignment and cooperative always proceed in discrete time. Meanwhile, the sub tasks of every plants are accomplished in continuous time. For instant, in distributed generation, dispatch interval of renewable energy resources is at an hourly timescale, but the distributed generators are operated in continuous time \cite{Gangammanavar}.  The research of stabilization for this kind of hybrid system is long-standing. In the classic work \cite{Kalman} presented by Kalman, et.al., the physical plants are modelled by continuous differential equations, and the control signal is generated by a hybrid system with continuous part and discrete part. Thus, in this work we study the sampled data cooperative output regulation of linear multi-agent systems with a hybrid control approach. This hybrid control method is implemented by a sampled data compensator.

The remainder of the article is organised as follows. In Section 2, some preliminaries are given, and the
problem to be solved is formulated. In Section 3, a distributed sampled data feedback controller is
proposed, and conditions for the cooperative output regulation are derived. Simulation examples are given in Section 4, concluding a numerical simulation example and an application example of micro-grid control, and finally, Section 5 concludes the article.

\section{Mathematic preliminaries and problem formulation}

\subsection{Mathematic preliminaries}
First of all, some basic mathematic notations should be presented. $\mathbb{C}$ denotes the set of complex numbers, $\mathbb{C}^+$ denotes the set of complex numbers with positive or zero real parts, i.e. $\mathbb{C}^+:=\{\lambda \in C|Re(\lambda)\geq 0\}$, conversely, $\mathbb{C}^-$ denotes the set of complex numbers with negative real parts, i.e. $\mathbb{C}^-:=\{\lambda \in C|Re(\lambda) < 0\}$. $\mathbb{R}$ denotes the set of real numbers, $\mathbb{R}^n$ denotes the n-dimensional Euclidean space. For $x \in {\mathbb{R}^n}$, and ${x^T}$ denotes its transpose. $\mathbb{R}^{m\times n}$ are the set of real matrices with $m$ rows and $n$ columns. $\mathbf{1}_N$ denotes a $N\times 1$ column vector whose elements are all 1, $\mathbf{0}_{m\times n}$ notes a zero matrix $m$ rows and $n$ columns and $I_{mn}$ notes a identity matrix with  $m\cdot n$ rows and columns.

For a matrix $A\in \mathbb{R}^{n\times n}$, $\lambda_{i}(A)$ is the $i$th eigenvalues of $A$ and $\sigma(A)$ denotes its spectrum. $\sigma_{max}(A)$ denotes the maximum spectrum of $A$. A square matrix $A$ is said to be $A\in \mathcal{D}_1$ if and only if $\sigma(A) < 1$. $\iota$ denote the imagine unit $\sqrt { - 1}$. For the complex number $\alpha \in \mathbb{C}$, its modulus is denoted as $| \alpha | $, $\Re(\alpha)$ and $\Im(\alpha)$ denotes the real part and the imaginary part of $\alpha$. For a real number series $a_i \in \mathbb{R} $ $(i = 1,\dots,n)$, $\min (a_i)$ denotes the minimum number of $a_i  $.

 $f(t^+)$ denotes $lim_{\varepsilon\rightarrow 0^+}f(t-\varepsilon)$ assuming that the right limit exists. $\mathbf{f}(\cdot)$ represent a function vector, and $f_i(\cdot)$ is the $ith$ component of the vector.

For $A= (a_{ij})\in R^{n\times m}$, $B= (b_{ij})\in R^{p\times q}$, the Kronecker product of $A$ and $B$ is defined as
\begin{equation*}
A \otimes B = \left( {\begin{array}{*{20}{c}}
{{a_{11}}B}& \cdots &{{a_{1m}}B}\\
 \vdots & \ddots & \vdots \\
{{a_{n1}}B}& \cdots &{{a_{nm}}B}
\end{array}} \right),
\end{equation*}
Another well-known results on Kronecker product should listed in the following.

Assuming the $A$ and $B$ are square matrices, with $n$ dimensions and $q$ dimensions respectively.  ${\lambda _i}(A),(i = 1,\dots,n)$ are the eigenvalues of $A$, $ {\lambda _j}(B),(j = 1,\dots,q)$ are the eigenvalues of $B$. The eigenvalue of $A\otimes B$ are ${\lambda _i}(A)\cdot {\lambda _j}(B),i = 1,\dots,n, j = 1,\dots,q$.

\subsection{Preliminaries of graph theory}
The graph theory has been presented many works about multi-agent system \cite{Saber2004}, but for the convenience, we should explain it as follows.
We consider the heterogeneous linear system consisting of $N+1$  systems with extra reference signal generated by an exosystem. The interaction topology of connection between $N+1$ agents is described by an undirected graph
$\mathcal {G}=(\mathcal {V},\mathcal {E})$ , where $\mathcal {V}=\{0,1,\ldots,N\}$ represent the node set (in this paper $0$ node represents the exosystem, and node $i=1,\dots,N$ note the agents), and $\mathcal {E}\subseteq\mathcal {V}\times\mathcal {V}$ notes the edge set. The weight adjacency matrix of $\mathcal {G}$ is denoted by $\mathcal{A}=[a_{ij}]\in \mathbb{R}^{(N+1)\times (N+1)}$, where $a_{ii}=0$ and $a_{ij}=a_{ji}\geq 0$ for $i\neq j$. Moreover, $a_{ij}=a_{ji}>0$ if and only if there exists an edge between node $i$ and node $j$. If there is an edge between two nodes, the two nodes are called neighbors. The set of neighbors of node $i$ is denoted by $\mathcal {N}_{i}=\{j\in \mathcal {V}: (i,j)\in \mathcal {E}, i\neq j\}$ . The degree matrix of $\mathcal {G}$ is $\mathcal {D}=diag(d_0,\ldots,d_{N})\in \mathbb{R}^{N+1\times N+1}$, where $d_i=\sum_{j=1}^{N}a_{ij}$ for $i=0,\ldots,N$. Then the Laplacian of the weighted graph is defined as $\mathcal {L}=\mathcal {D}-\mathcal {A}$.

In the network $\mathcal {G}=(\mathcal {V},\mathcal {E})$, the subset of nodes $\mathcal {\bar V} = \{1,\dots,N\}$ can be divided into two sets, one set of nodes which can access the exogenous signal $w$, we set $a_{i0}>0$, otherwise $a_{i0}=0$ for the set which can not access the exogenous signal.

Additionally, by removing all edges between the node $0$ and the nodes in $\mathcal{\bar V}$. The left edges and node set $\mathcal{\bar V}$ construct a subgraph $\mathcal {\bar G}=(\mathcal {\bar V},\mathcal {\bar E})$. We use $\mathcal{\bar L}$ denotes the Laplacian of $\mathcal {\bar G}$. For node $i=1,\dots,N$, the set of neighbors of node $i$ in the subgraph $\mathcal{\bar G}$ is denoted by $\mathcal{\bar N}_i$.

The Laplacian $\mathcal {L}$ of $\mathcal{G}$ can be partitioned as follows
\begin{equation*}
\mathcal{L} = \left( {\begin{array}{*{20}{c}}
{\sum\nolimits_{j = 1}^N {{a_{0j}}} }&\vline& {\left[ {{a_{01}}, \ldots ,{a_{0N}}} \right]}\\
\hline
{ - \Delta {\mathbf{1}_N}}&\vline& H
\end{array}} \right)
\end{equation*}
where $\Delta$ be an $N\times N$  diagonal matrix, with $i$th diagonal element is $a_{i0}$ for $i=1,\dots,N$.

Worth to mention, $H = \mathcal{\bar L} + \Delta$ and $H\cdot \mathbf{1}_N =\Delta\cdot \mathbf{1}_N$, since $\mathcal{L}\mathbf{1}_{N}= 0$. These results and the following Lemma can be found in \cite{huangjie2012}.
\begin{lemmax}
All the nonzero eigenvalues of $H$, if any, have positive real parts. Furthermore, $H$ is nonsingular if and only if the digraph $\mathcal{G}$ contains a directed spanning tree with the node $0$ as its root.
\end{lemmax}

For Lemma 1, some explanation should be given. If the graph $\mathcal{G}$ contains an sequence of edges in this form $(0, i_1), (i_1, i_2),\dots, (i_{l-1}, i_l)$, then the set $(0, i_1), (i_1, i_2),\dots, (i_{l-1}, i_l)$ is called a path of $\mathcal{G}$ from node $0$ to node $i_l$, and $i_l$ is reachable from node $0$. If each node $i\in \mathcal{\bar V}$ has exactly one parent and is reachable from $0$, moreover $0$ has no parents. Then the digraph $\mathcal{G}$ contains a directed spanning tree with the node $0$ as its root.\\\\
Another important Lemma should given are as follows, which are rephrased from Lemma 1 of \cite{Teel}.
\begin{lemmax}
Considering the cascade of two hybrid linear systems as follows
\begin{equation}\label{1}
\begin{array}{l}
\left\{ \begin{array}{l}
\dot z(t) = F(t)z(t) + G(t)w(t)\\
\dot w(t) = Sw(t)
\end{array} \right.,\quad t \in ({t_k},{t_{k + 1}}]\\
\left\{ \begin{array}{l}
 z({t_k^+}) = Mz(t_k) + Nw(t_k)\\
 w({t_k^+}) = Jw(t_k)
\end{array} \right.,
\end{array}
\end{equation}
where $F(t)$ and $G(t)$ are continuous matrix-valued functions, $z(t)\in \mathbb{R}^n$, $w(t)\in \mathbb{R}^s$, $z(t_k^+)\in \mathbb{R}^n$ and $w(t_k^+)\in \mathbb{R}^s$.   The interval of jump period $h:=(t_{k+1}-t_k) >0$, and $h:=(t_{k+1}-t_k) >0$, $\mathop {\lim }\limits_{k \to  + \infty } {t_k} =  + \infty $. And $\phi(h)$ is defined as $\phi(h)=\exp{(F\cdot h)}$.

Assume that
\begin{equation}\label{2}
\sigma (M\phi(h))\cap \sigma (J \exp(S\cdot h))=\emptyset
\end{equation}
then there exist a unique continuously differentiable solution  $\Pi (t):({t_k},{t_{k + 1}}] \to {\mathbb{R}^{n \times s}}$
to the equations
\begin{equation}\label{3}
\frac{{d\Pi (t)}}{{dt}} = F(t)\Pi (t) - \Pi (t)S + G,\quad t\in ({t_k},{t_{k + 1}}]
\end{equation}
and the $\Pi ({t_k^+})$ satisfied
\begin{equation}\label{pi}
0 = M\Pi ({t_{k }}) - \Pi ({t_k^+})J + N
\end{equation}
as a consequence, the set $\mathcal{M}_1=\{ z|z(t)=\Pi(t)w(t)\}$ is invariant for (\ref{1}), Furthermore, if $M\phi(h)\in \mathcal{D}_1$ then the set $\mathcal{M}$ in (\ref{1}) is globally exponentially stable for  $\mathcal{M}_1$.
\end{lemmax}


\subsection{Problem formulation}
 In this technical note, we will further study the cooperative continuous time heterogeneous linear multi-agent systems with partial states consensus with sample compensator of the following form. Consider $N$ nonidentical state-space models
\begin{equation}\label{5}
\begin{array}{l}
{{\dot x}_i} = {A_i}{x_i} + {B_i}{u_i}+Pw_i\\
{e_i} = C_i{x_i}+Q_iw,\quad\quad\qquad \qquad \qquad i=1,\dots,N
\end{array}
\end{equation}
where ${x_i} \in {\mathbb{R}^{{n_i}}}$, ${e_i} \in {\mathbb{R}^{{p_i}}}$ , and ${u_i} \in {\mathbb{R}^{{m_i}}}$ are the state vectors, error vectors, and input vectors of the $i$th subsystem.
And  $w  \in {\mathbb{R}^q}$ is the exogenous signal representing the reference input to be tracked which generated by an exosystem with the dynamic as follows
\begin{equation}\label{6}
\dot w = Sw
\end{equation}

The objective here is to design a distributed control law such that for all the agents $i\in \mathcal{\bar V}$ the tracking error approaches to zero asymptotically. Additionally, the a distributed control law using only sampled data from $i$th agents  states and their neighbors, and the sampled data from exosystem for some agents can access the information of reference system (\ref{6}).  This problem with continuous control law has been studied in \cite{huangjie2012}, and was called cooperative output regulation problem.

As mentioned in section 2.2, the $i$th agents ($i=1,\dots,N$) in (\ref{5}) can be classified into two sets of agents. One set of agents consists of those agents whose controller can access the reference signal, and the other set of agents consists of the rest of the  agents which cannot access the exogenous signal as a part of the input. Then we introduce our distributed control law with sampled data as follows.


As in \cite{Yholdfun}, $H_i(t)$ is a fixed $h$-periodic hold function which defined in terms of a pair of matrix $C_{Hi}$ and $A_{Hi}$, via
\begin{equation*}
H_i(t)=C_{Hi}\exp (A_{Hi}(t)), \quad t\in (t_k,t_{k+1}],
\end{equation*}
and
\begin{equation*}
{u_i} = {H_i}(t){u_i}(k),
\end{equation*}
Worth to mention that, $H_i(t)$, $C_{Hi}$ and $A_{Hi}$ has no strictly requirement on dimension, they only need to make sure $u_i \in {\mathbb{R}^{{m_i}}}$ and $A_{Hi}$ is nonsingular square matrix.

We introduce intermediate variable $\xi_i(t)$ to express the h-periodic hold control signal, then the sampled data controller of agent $i$ can be represented as the following hybrid distribute controller
\begin{equation}\label{ithcloseloop}
 \begin{array}{l}
{{\dot x}_i}(t) = {A_i}x_i(t) + B_iC_{Hi}\xi_i(t),\quad t \in ({t_k},{t_{k + 1}}]\\
\dot\xi_i(t)=A_{Hi}\xi_i(t),\quad t \in ({t_k},{t_{k + 1}}]\\
{\xi_i}(t_k^+)  = f_i({\xi_i}(t_k), {\eta _i}(t_k),x_i(t_k))\\
\dot{\eta _i}(t) = S{\eta _i}(t),\quad t \in ({t_k},{t_{k + 1}}]\\
\eta _i(t_k^+)  = \eta _i(t_k) - {\mu}\left(\sum\limits_{j \in {{\cal N}_i}} {{a_{ij}}({\eta _j}(t_k ) - {\eta _i}(t_k ))}+ a_{i0}(w(t_k ) - {\eta _i}(t_k )) \right) \\
\dot{w}=Sw\\
{e_i} = {C_i}{x_i} + {Q_i}{w _i}
\end{array} \quad i = 1,2,...,N,
\end{equation}
where ${\eta _i} \in {\mathbb{R}^q}$, and ${\mu}$ is some positive number.

Then we proposed the following definition which derive from the definition 1 in \cite{huangjie2012}.
\begin{definition}Cooperative Output Regulation Problem (CORP) with sample compensator:\\
Systems (\ref{5}) and System (\ref{6}) is said to achieve cooperative output regulation with compensator (\ref{ithcloseloop}) if the following condition holds:
For any initial condition $x_i(0)$, $\eta_i(0)$, and $w(0)$, the tracking error
\begin{equation}
 \mathop {lim}\limits_{t \to \infty } {e_i}(t)  = 0, \quad  i \in 1,2,...,N.
\end{equation}
\end{definition}

\section{Solvability of the problem}
Before discuss the solution of CORP, some basic assumptions of the sampled data controller for this problem should be list in following lines.

\begin{assumption}
$S$ has no eigenvalues with negative real parts, i.e. $\sigma(S)\in  \mathbb{C}^+$.
\end{assumption}
\begin{assumption}
the digraph $\mathcal{ G}$ contains a directed spanning tree with the node $0$ as the root.
\end{assumption}
Without of generality, we assume $f_i(x_i,\eta_i)$ in (\ref{ithcloseloop}) has the following form
\begin{equation*}
f_i({\xi_i}(t_k), {\eta _i}(t_k),x_i(t_k))= {K_{1i}}x_i(t_k)+K_{2i}\eta_i(t_k),
\end{equation*}
where  ${K_{1i}} \in {\mathbb{R}^{{m_i} \times {n_i}}}$, and ${K_{2i}} \in {\mathbb{R}^{{m_i} \times {q_i}}}$ are gain matrices to be determined. \\
For the $i$th agent ($i\in \mathcal{\bar V}$), letting $z_i(t)=[x_i^T(t),\xi_i^T(t)]^T$ when $t\in (t_k,t_{k+1}]$, and $z_i(t_k)=[x_i^T(t_k),\xi_i^T(t_k)]^T$.\\\\
Then, we note $F_i=\left[ {\begin{array}{*{20}{c}}
{{A_i}}&{{B_i}{C_{Hi}}}\\
0_{}&{{A_{Hi}}}
\end{array}} \right]$ and  $G_i=\left[ {\begin{array}{*{20}{c}}
{{P_i}}\\
0
\end{array}} \right]$,${M_i} = \left[ {\begin{array}{*{20}{c}}
I_{ni}&0\\
{{K_{1i}}} & 0
\end{array}} \right]$, ${\Gamma_i} = \left[ {\begin{array}{*{20}{c}}
0\\
{{K_{2i}}}
\end{array}} \right]$,
and ${\hat C_i} = \left[ {\begin{array}{*{20}{c}}
{{C_i}}&0
\end{array}} \right]$ (where $\hat C_i \in \mathbb{R}^{(n_i+q)\times n_i}$), for $i = 1,\dots N$.\\\\
If the interior structure of above matrices are not taken in to consideration, we have a generalized hybrid multi-agents system as
\begin{equation}\label{hccorp}
 \left( {\begin{array}{*{20}{c}}
{{{\dot z}_i}(t)}\\
{{{\dot \eta }_i}(t)}
\end{array}} \right) = \left[ {\begin{array}{*{20}{c}}
{{F_i}}&0\\
0&S
\end{array}} \right]\left( {\begin{array}{*{20}{c}}
{{z_i}(t)}\\
{{\eta _i}(t)}
\end{array}} \right) + \left[ {\begin{array}{*{20}{c}}
{{G_i}}\\
0
\end{array}} \right]w(t),\quad t\in (t_k,t_{k+1}]
\end{equation}

and
\begin{equation}\label{hdcorp}
\left( {\begin{array}{*{20}{c}}
{{z_i}({t_k^+})}\\
{{\eta _i}({t_k^+})}
\end{array}} \right) = \left[ {\begin{array}{*{20}{c}}
{{M_i}}&{{\Gamma_i}}\\
0&I
\end{array}} \right]\left( {\begin{array}{*{20}{c}}
{{z_i}(t_k)}\\
{{\eta _i}(t_k)}
\end{array}} \right) + \left[ {\begin{array}{*{20}{c}}
0\\
\mu
\end{array}} \right]\left( {\sum\limits_{j \in {N_i}} {{a_{ij}}({\eta _j}(t_k ) - {\eta _i}(t_k ))}  + {a_{i0}}(w(t_k ) - {\eta _i}(t_k ))} \right).
\end{equation}

Then let $\widetilde{F}={\rm blockdiag}(F_1,\dots,F_n)$,
$\widetilde{G} ={\rm blockdiag}(G_1,\dots,G_n)$, $\widetilde{S} =I_N\otimes S$, $\widetilde{M}={\rm blockdiag}(M_1,\dots,M_n)$, \\
$\widetilde{\Gamma}={\rm blockdiag}(\Gamma_1,\dots,\Gamma_n)$.

Let $z = [z_1^T, \ldots ,z_N^T]$, $\eta  = [\eta _1^T, \ldots ,\eta _N^T]$, and $\tilde w=1_N\otimes w$.
Thus, the multi-agent under control can be separated in to continuous time system (\ref{networkc}) when $t\in(t_k,t_{k+1}]$, and discrete time system (\ref{networkd}) as follows
\begin{equation}\label{networkc}
\left( {\begin{array}{*{20}{c}}
{\dot z(t)}\\
{\dot \eta (t)}
\end{array}} \right) = \left[ {\begin{array}{*{20}{c}}
{\tilde F}&0\\
0&{\tilde S}
\end{array}} \right]\left( {\begin{array}{*{20}{c}}
{z(t)}\\
{\eta (t)}
\end{array}} \right) + \left[ {\begin{array}{*{20}{c}}
{\tilde G}\\
0
\end{array}} \right]\tilde w(t),\quad t\in (t_k,t_{k+1}],
\end{equation}
and
\begin{equation}\label{networkd}
\left( {\begin{array}{*{20}{c}}
{z({t_k})}\\
{\eta ({t_k})}
\end{array}} \right) = \left[ {\begin{array}{*{20}{c}}
{\tilde M}&{{{\tilde \Gamma}}}\\
0&{I_{qN} - \mu (H \otimes {I_q})}
\end{array}} \right]\left( {\begin{array}{*{20}{c}}
{z(t_k)}\\
{\eta (t_k)}
\end{array}} \right) + \left[ {\begin{array}{*{20}{c}}
0\\
{\mu (H \otimes {I_q})}
\end{array}} \right]\tilde w(t_k).
\end{equation}\\
Before we introduce the detailed designing method of sampled data compensator, we carry out the cooperative output regulation condition of hybrid system above ((\ref{networkc}) and (\ref{networkd}))  though theorem 1.
\begin{theoremx}
Under Assumptions 1 and 2, if $M_i\cdot\exp(F_i\cdot h)\in \mathcal{D}_1$, additionally, for each plant, the continuous matrix-value functions $\Pi_i(t)$ satisfying
\begin{equation}\label{11}
0 = \hat C_i \Pi_i (t) - Q_i,\quad i=1,\dots,N.
\end{equation}

where $\Pi_i(t)$ defined as
\begin{equation*}
\Pi_i(t)=\{\exp(F_i\cdot t)\Pi_i(t_k)+{\int_0^t  {\exp [{F_i}(t  - \theta )] \cdot {G_i} \cdot \exp (S\cdot \theta )d\theta } }\}\exp(-S\cdot t).
\end{equation*}

Then the CORP for (\ref{hccorp}) and (\ref{hdcorp}) can be solved by choosing appropriate positive real number $\mu$ and $h$ such that
\begin{equation}
0<\mu<\frac{{4\cdot{{\min(\Re(\lambda(H)))}}}}{\sigma^2_{max}(H)}.
\end{equation}
\end{theoremx}
\noindent \textbf{Proof.}\,\,
We separate the proof in to two parts, the regulation requirement (RF) condition part and the stability requirement (SF) condition part.
RF is corresponding to equation (\ref{3}) and (\ref{pi}) is established in lemma 2 for the networked hybrid system (\ref{networkc}) and (\ref{networkd}).
SF is corresponding to the requirement of $M\phi(h)\in \mathcal{D}_1$  in lemma 2.

Firstly, we verify condition RF is full filled.

For each pair $F_i$, $G_i$ ($i\in \mathcal{\bar V}$) there exists an continuously differentiable solution to the equations
\begin{equation}
\frac{{d{\Pi _i}(t)}}{{dt}} = {F_i}{\Pi _i}(t) - \Pi_i (t)S + {G_i}, t\in(t_k,t_{k+1}],
\end{equation}
and $\Pi_i(t)=[\exp(F_i\cdot t)\Pi_i(t_k)+L_i(\tau)]\exp(-S\cdot\tau)$, where $L_i(t)= {\int_0^t  {\exp [{F_i}(t  - \theta )] \cdot {G_i} \cdot \exp (S \theta )d\theta } }$.

We note $\Pi(t)= {\rm blockdiag} (\Pi_1(t),\dots,\Pi_N(t))$ and let $\tilde \Pi(t)= \left( {\begin{array}{*{20}{c}}
\Pi(t) \\
{{I_{q  N}}}
\end{array}}\right)$. Then the following equation (\ref{networkpi}) is established
\begin{equation}\label{networkpi}
\frac{{d\tilde \Pi }}{{dt}} = \left[ {\begin{array}{*{20}{c}}
{\tilde F}&0\\
0&{\tilde S}
\end{array}} \right]\tilde \Pi  - \tilde \Pi \left[ {\begin{array}{*{20}{c}}
{\tilde S}&0\\
0&{\tilde S}
\end{array}} \right] + \left[ {\begin{array}{*{20}{c}}
{\tilde G}\\
0
\end{array}} \right].
\end{equation}
\\

With appropriately designed  $M_i$ and $h$ choosing rightly, we can make sure that $M_i\exp(F_i\cdot h)\in \mathcal{D}_1$, and according to assumption 1, we have $\exp(S\cdot h)  \notin \mathcal{D}_1$,
then there exist an unique ${\Pi _i}({t_k^+})$ which satisfied the following equation
\begin{equation}\label{15}
{M_i}\exp ({F_i}h){\Pi _i}({t_k^+}) - {\Pi _i}({t_k^+})\exp (Sh) + {\Gamma_{i}}\exp (Sh)  + {M_i}{L_i}(h),
\end{equation}
and equation (\ref{16}) is obviously established.
\begin{equation}\label{16}
({I_{q  N}} - \mu ({H_i} \otimes {I_q}))\exp(\tilde Sh){{I}_{q  N}}- {{I}_{q  N}}\exp(\tilde Sh) +  \mu (H \otimes {I_q})\exp(\tilde Sh) = 0.
\end{equation}
With equation (\ref{15}) and (\ref{16}), we can get (\ref{17}),
\begin{multline}\label{17}
\left[ {\begin{array}{*{20}{c}}
{\tilde M}&{{{\tilde \Gamma}}}\\
0&{I_{qN} - \mu (H \otimes {I_q})}
\end{array}} \right]\exp \left( {\left[ {\begin{array}{*{20}{c}}
{\tilde F}&0\\
0&{\tilde S}
\end{array}} \right] \cdot h} \right)\tilde \Pi ({t_k}) - \tilde \Pi ({t_k})\exp \left( {\left[ {\begin{array}{*{20}{c}}
{\tilde F}&0\\
0&{\tilde S}
\end{array}} \right] \cdot h} \right) \\
  + \left[ {\begin{array}{*{20}{c}}
0\\
{I_{qN} -\mu (H \otimes {I_q})}
\end{array}} \right]\exp \left( {\left[ {\begin{array}{*{20}{c}}
{\tilde S}&0\\
0&{\tilde S}
\end{array}} \right] \cdot h} \right) + \left[ {\begin{array}{*{20}{c}}
{\tilde M}&{{{\tilde \Gamma}}}\\
0&{ \mu (H \otimes {I_q})}
\end{array}} \right]\tilde L(h)
\end{multline}
where $\tilde L(h) = {[\begin{array}{*{20}{c}}
{{L^T(h)}}&0
\end{array}]^T}$, in which $L(h)=blockdiag(L_1(h),\dots,L_N(h))$.

Then we have the RF condition is full filled for networked hybrid system (\ref{networkc}) and (\ref{networkd}).

Secondly, we check the SF condition is full filled.

For each agent, if $M_i\cdot \exp(F_i\cdot h)\in \mathcal{D}_i$ is satisfied with appropriately designed $M_i$ and $h$ chose wisely, then we have $\tilde M \cdot \exp(\tilde F\cdot h)\in \mathcal{D}_i$.

We note $\mathbf{M}_1 = [I_{qN}-\mu (H \otimes {I_q})] \cdot \exp(\tilde S\cdot h)$, then we have
\begin{eqnarray}\label{20}
                {\mathbf{M}_1} &=& [{I_{qN}} - \mu (H \otimes {I_q})] \cdot \exp (\tilde S \cdot h) \nonumber\\
                 &= &[{I_{qN}} - \mu (H \otimes {I_q})] \cdot [{I_N} \otimes \exp (S \cdot h)] \nonumber\\
                 &= &[({I_N} - \mu H) \otimes {I_q}] \cdot [{I_N} \otimes \exp (S \cdot h)] \nonumber\\
                 &= &({I_N} - \mu H) \otimes \exp (S \cdot h)
\end{eqnarray}
For the matrix $\mathbf{M}_1$, each eigenvalue $\lambda_l(\mathbf{M}_1) = [1-\mu\lambda_i(H)] \cdot \lambda_{v}(\exp(S\cdot h)(i=1,\dots,N, v=1,\dots q)$, we have
$\sigma_l(\mathbf{M}_1) \leq \sigma_i[I-\mu(H)] \cdot \sigma_{v}(\exp(S\cdot h)(i=1,\dots,N, v=1,\dots q)$.

Without loss of generality, we assume the $\lambda_i(H) = a_i + b_i \iota, (a_i , b_i \in \mathbb{R})$. Then we have
$\sigma_i[I-\mu(H)] = \sqrt {{{(1 - \mu {a_i})}^2} + {{(\mu {b_i})}^2}} $. With Assumption 1 satisfied,  we have ${\sigma _v^2(\exp (S \cdot h)} \geq 1$.

By using Lemma 1, with the Assumption 2 satisfied, each eigenvalue of $H$ has a positive real part, i.e. $a>0$.

Letting $c_v = {\sigma _v^2(\exp (S \cdot h)}$, for $i=1,\dots,N, v=1,\dots,q$, we have the following result
\begin{eqnarray}\label{21}
  {\sigma _l}({\mathbf{M}_1}) < 1  &\Leftarrow& {\sigma _i}[I - \mu (H)] \cdot {\sigma _v}(\exp (S \cdot h) < 1 \Leftrightarrow {\sigma^2 _i}[I - \mu (H)] \cdot {\sigma^2 _v}(\exp (S \cdot h) < 1 \nonumber \\
   &\Leftrightarrow& [{(1 - \mu {a_i})^2} + {(\mu {b_i})^2}] \cdot \sigma _v^2(\exp (S \cdot h) < 1 \nonumber\\
   &\Leftrightarrow& [{(1 - \mu {a_i})^2} + {(\mu {b_i})^2}] < \frac{1}{{\sigma _v^2(\exp (S \cdot h)}} \nonumber\\
   &\Leftrightarrow& \frac{{2a - \sqrt {4a_i^2 - 4(a_i^2 + b_i^2)(1 - {c_v})} }}{{(a_i^2 + b_i^2)}} < \mu  < \frac{{2a + \sqrt {4a_i^2 - 4(a_i^2 + b_i^2)(1 - {c_v})} }}{{(a_i^2 + b_i^2)}} \nonumber\\
   &\Leftarrow&  \frac{{2a - \sqrt {4a_i^2} }}{{(a_i^2 + b_i^2)}} < \mu  < \frac{{2a + \sqrt {4a_i^2} }}{{(a_i^2 + b_i^2)}} \Leftarrow 0 < \mu  < \frac{{4{a_i}}}{{(a_i^2 + b_i^2)}}\Leftarrow 0 < \mu  < \frac{{4{{\min(a_i)}}}}{\sigma^2_{max}(H)}.
\end{eqnarray}

With $0<\mu<\frac{{4\cdot{{\min(\Re(\lambda(H)))}}}}{\sigma^2_{max}(H)}$, then for any value of $h$, we have ${\sigma _i}[I - \mu (H)] \cdot {\sigma _v}(\exp (S \cdot h) < 1$, i.e., $\mathbf{M}_1 = ({I_{qN} -\mu (H \otimes {I_q})})\cdot \exp(\tilde S\cdot h)$ is Shur stable.

Therefore, with $\tilde M \cdot \exp(\tilde F\cdot h)\in \mathcal{D}_i$ and $({I_{qN} -\mu (H \otimes {I_q})})\cdot \exp(\tilde S\cdot h)$ is Shur stable, the following result is established
\begin{equation}\label{18}
\left[ {\begin{array}{*{20}{c}}
{\tilde M}&{\tilde \Gamma}\\
0&{I - \mu (H \otimes {I_q})}
\end{array}} \right]\exp \left( {\left[ {\begin{array}{*{20}{c}}
{\tilde F}&0\\
0&{\tilde S}
\end{array}} \right] \cdot h} \right) \in \mathcal{D}_1
\end{equation}

Then we can conclude that the SF condition is full filled.

According to (\ref{17}) and (\ref{18}), and Lemma 2, with RF and SF condition both full filled, then there exist an invariant set $\mathcal{M}=\{z|z(t)=\tilde \Pi(t) \tilde w(t)\}$ for
system (\ref{networkc}) and (\ref{networkd}), and it is is globally exponentially stable for system (\ref{networkc}) and (\ref{networkd}).

Considering (\ref{11}) is established, then we have
\begin{equation*}
  \mathop {lim}\limits_{t \to \infty } {e_i}(t)  = 0, \quad  i \in 1,2,...,N.
\end{equation*}
The proof is completed.

\subsection{Sampled data cooperative controller}
In this section, we discuss the sampled data control for CORP with (\ref{ithcloseloop}) in detail. Firstly, some essential assumptions for sampled data control should be introduced.
\begin{assumption}
For each plant, the pairs $({A_i},{B_i})$  are controllable, and sampling frequency is non-pathological, (i.e. $A_i$ and $S$ do not have two eigenvalue with equal real parts and imaginary parts that coincide with the $a\cdot h, a\in \mathbb{Z}$)\cite{Chenbook}.
\end{assumption}
\begin{assumption}
For each plant $i$, all $\lambda\in\sigma(S)$, the following equation is established
\begin{eqnarray*}
rank\left( {\begin{array}{*{20}{c}}
{A_i - \lambda I}&B_i\\
C_i&Q_i
\end{array}} \right) = n + p\quad.
\end{eqnarray*}
\end{assumption}
For each plant (\ref{ithcloseloop}), we choose $A_{Hi} = 0$ and $C_{Hi} = I_{ni}$, then we have the distribute zero-order hold $H_i(t) = I_{m} $ sampled data controller as follows
\begin{equation}\label{ithcloseloop2}
\begin{array}{*{20}{l}}
{{{\dot x}_i}(t) = {A_i}{x_i}(t) + {B_i}{\xi _i}(t)+Pw,\quad t \in ({t_k},{t_{k + 1}}]}\\
{{\xi _i}(t) = {\xi _i}(t_k^ + ),\quad t \in ({t_k},{t_{k + 1}}]}\\
{{\xi _i}(t_k^ + ) =  {K_{1i}}{x_i}({t_k})}+{K_{2i}}{\eta _i}({t_k})\\
{\mathop {{\eta _i}}\limits^. (t) = S{\eta _i}(t),\quad t \in ({t_k},{t_{k + 1}}]}\\
{{\eta _i}(t_k^ + ) = {\eta _i}({t_k}) - \mu \left( {\sum\limits_{j \in {N_i}} {{a_{ij}}({\eta _j}({t_k}) - {\eta _i}({t_k}))}  + {a_{i0}}(w({t_k}) - {\eta _i}({t_k}))} \right)}\\
{\dot w = Sw}\\
{{e_i} = {C_i}{x_i} + {Q_i}{w_i}}
\end{array}\quad i = 1,2,...,N,
\end{equation}
where $K_{1i} $ and $K_{2i} $ are gain matrices to be determined.

Considering the $ith$ plant under control (\ref{ithcloseloop2}), can be divided into the continuous part and the discontinuous part, for the continuous part, we have
\begin{equation}\label{sampledc}
\left[ {\begin{array}{*{20}{c}}
{{x_i}(t)}\\
{{\xi _i}(t)}\\
{{\eta _i}(t)}
\end{array}} \right] = \left[ {\begin{array}{*{20}{c}}
{{A_{Di}}(t) + {B_{Di}}(t){K_{1i}}}&0&{{B_{Di}}(t){K_{2i}}}\\
{{K_{1i}}}&0&{{K_{2i}}}\\
0&0&{{S_{Di}}(t)}
\end{array}} \right]\left[ {\begin{array}{*{20}{c}}
{x_i(t_k^ + )}\\
{{\xi _i}(t_k^ + )}\\
{\eta_i (t_k^ + )}
\end{array}} \right] + \left[ {\begin{array}{*{20}{c}}
{{P_{Di}}(t)}\\
0\\
0
\end{array}} \right]w(t_k^ + ),\quad t \in ({t_k},{t_{k + 1}}],
\end{equation}
and the discontinuous part
\begin{equation}\label{sampledd}
\begin{array}{l}
\left[ {\begin{array}{*{20}{c}}
{{x_i}(t_{k }^ + )}\\
{{\xi _i}(t_{k }^ + )}\\
{{\eta _i}(t_{k }^ + )}
\end{array}} \right] = \left[ {\begin{array}{*{20}{c}}
I&0&0\\
{{K_{1i}}}&0&{{K_{2i}}}\\
0&0&I
\end{array}} \right]\left[ {\begin{array}{*{20}{c}}
{{x_i}({t_{k }})}\\
{{\xi _i}({t_{k }})}\\
{{\eta _i}({t_{k }})}
\end{array}} \right]\\
\quad \quad \quad \quad \quad  + \left[ {\begin{array}{*{20}{c}}
\begin{array}{l}
0\\
0
\end{array}\\
\mu
\end{array}} \right]\left( {\sum\limits_{j \in {N_i}} {{a_{ij}}({\eta _j}({t_{k }}) - {\eta _i}({t_{k }}))}  + {a_{i0}}(w({t_{k }}) - {\eta _i}({t_{k }}))} \right).
\end{array}
\end{equation}
where $A_{Di}(t)=\exp\left({A\cdot t}\right)$, ${B_{Di}}(t) = \int_{{t_k}}^t {\exp \left( {{A_i} \cdot (t - s)} \right)}{B_i}ds$, ${S_{Di}}(t) = \exp \left( {S \cdot t} \right)$ and \\
$P_{Di}=\int_{{t_k}}^t {\exp \left( {{A_i} \cdot (t - s)} \right)}{P_i}\exp(S\cdot s)ds$. \\
Moreover if $t=h$, we have
\begin{equation}\label{sampledh}
\left[ {\begin{array}{*{20}{c}}
{{x_i}({t_{k + 1}})}\\
{{\xi _i}({t_{k + 1}})}\\
{{\eta _i}({t_{k + 1}})}
\end{array}} \right] = \left[ {\begin{array}{*{20}{c}}
{{A_{Di}}(h) + {B_{Di}}(h){K_{1i}}}&0&{{B_{Di}}(h){K_{2i}}}\\
{{K_{1i}}}&0&{{K_{2i}}}\\
0&0&{{S_{Di}}(t)}
\end{array}} \right]\left[ {\begin{array}{*{20}{c}}
{x_i(t_k^ + )}\\
{{\xi _i}(t_k^ + )}\\
{\eta_i (t_k^ + )}
\end{array}} \right] + \left[ {\begin{array}{*{20}{c}}
{{P_{Di}}(h)}\\
0\\
0
\end{array}} \right]w(t_k^ + ),\quad t \in ({t_k},{t_{k + 1}}],
\end{equation}

According to Theorem 3.2.2 in \cite{Chenbook}, under Assumption 3, for each agents $i$ ($i\in1,\dots,N$), $(A_{Di}(h), B_{Di}(h))$ are controllable.
Then there exist $K_{1i}$ such that ${{A_{Di}}(h) + {B_{Di}}(h){K_{1i}}}$ are Shur stable, let $K_{2i}$ be as follows
\begin{equation}\label{control2}
K_{2i} =  - K_{1i}\Pi _{i},
\end{equation}
where  $\Pi _{i}$ are decided by following equation
\begin{equation}\label{pii}
\begin{array}{l}
{\Pi_i }S = {A_i}{\Pi_i } +P_i\\
0 = {C_{i}}{\Pi _{i}} + {Q_{i}}
\end{array}
\end{equation}
Then we proposed the following theorem for the sampled data CORP.
\begin{theoremx}
Under Assumptions 1 to 4, the CORP can be solved by the sampled distributed feedback control law (\ref{ithcloseloop2}) with $K_{1i}, K_{2i}$ being such that (\ref{sampledh}) is Schur stable and (\ref{control2}) is established, and with the real number $\mu$ such that
\begin{equation}\label{29}
0<\mu<\frac{{4\cdot{{\min(\Re(\lambda(H)))}}}}{\sigma^2_{max}(H)}.
\end{equation}
\end{theoremx}
\noindent \textbf{Proof.}\,\,\\
We separate the proof in to two parts, the regulation requirement (RF) part and the stability requirement (SF) part.
RF is corresponding to equation (\ref{3}) and (\ref{pi}) in lemma 2  for the networked multi-agent system (\ref{ithcloseloop2}),
SF is corresponding to the requirement of $M\phi(h)\in \mathcal{D}_1$.

Firstly, let us verify that requirement (RF) is fulled.

To this aim, let $\widetilde{A}={\rm blockdiag}(A_{1},\dots,A_{N})$,
$\widetilde{B} ={\rm blockdiag} (B_{1},\dots,B_{N})$, $\tilde P = {\rm blockdiag} (P_{1},\dots,P_{N})$ and $\widetilde{S} =I_N\otimes {S}$.
Additionally, let feedback $\widetilde{K}_{1}={\rm blockdiag}(K_{11},\dots,K_{1N})$, $\widetilde{K}_{2}={\rm blockdiag}(K_{21},\dots,K_{2N})$ and states  $x(t) = [x_1^T(t), \ldots ,x_N^T(t)]$, $\eta(t)  = [\eta _1^T(t), \ldots ,\eta _N^T(t)]$, and $ \tilde w(t)=\mathbf{1}_N\otimes w(t)$.
Then from (\ref{ithcloseloop2}) we have
\begin{equation}\label{snetworkc}
\left[ {\begin{array}{*{20}{c}}
{\dot x(t)}\\
{\dot \xi (t)}\\
{\dot \eta (t)}
\end{array}} \right] = \left[ {\begin{array}{*{20}{c}}
{\tilde A}&{\tilde B}&0\\
0&0&0\\
0&0&{{{\tilde S}_{Di}}(t)}
\end{array}} \right]\left[ {\begin{array}{*{20}{c}}
{x(t)}\\
{\xi (t)}\\
{\eta (t)}
\end{array}} \right] + \left[ {\begin{array}{*{20}{c}}
{\tilde P}\\
0\\
0
\end{array}} \right]\tilde w(t),\quad t \in ({t_k},{t_{k + 1}}],
\end{equation}
and similarly we have
\begin{equation}\label{snetworkd}
\left[ {\begin{array}{*{20}{c}}
{x(t_{k }^ + )}\\
{\xi (t_{k }^ + )}\\
{\eta (t_{k }^ + )}
\end{array}} \right] = \left[ {\begin{array}{*{20}{c}}
I&0&0\\
{{\tilde K_1}}&0& \tilde K_2\\
0&0&{I - \mu (H \otimes {I_q})}
\end{array}} \right]\left[ {\begin{array}{*{20}{c}}
{x({t_{k }})}\\
{\xi ({t_{k }})}\\
{\eta ({t_{k }})}
\end{array}} \right] + \left[ {\begin{array}{*{20}{c}}
\begin{array}{l}
0\\
0
\end{array}\\
{\mu (\Delta \otimes {I_q})}
\end{array}} \right]\tilde w({t_k}).
\end{equation}
With Assumption 4, for each agents there exist a unique $\Pi_i$ satisfied equation (\ref{pii}) (Theorem 1.9 in \cite{Huang2004}).\\
Let $\Pi= {\rm blockdiag} (\Pi_1,\dots,\Pi_N)$ and then let $\tilde \Pi  = \left( {\begin{array}{*{20}{c}}
\Pi \\
0\\
{{I_{qN}}}
\end{array}} \right)$. Thus, from (\ref{snetworkc}) the following two equations is established
\begin{equation*}
 \tilde \Pi  \cdot \tilde S = \left[ {\begin{array}{*{20}{c}}
{\tilde A}&{\tilde B}&0\\
0&0&0\\
0&0&{\tilde S}
\end{array}} \right] \cdot \tilde \Pi  + \left[ {\begin{array}{*{20}{c}}
{\tilde P}\\
0\\
0
\end{array}} \right],\quad t \in ({t_k},{t_{k + 1}}],
\end{equation*}
and with $(\Delta \otimes I_q)(\mathbf{1}_N \otimes  w) = (H \otimes I_q)( \mathbf{1}_N \otimes  w)$ from (\ref{snetworkd}) we have
\begin{equation*}
 0 = \left[ {\begin{array}{*{20}{c}}
I&0&0\\
{{\tilde K_1}}&0&{ - {{\tilde K}_1}\Pi }\\
0&0&{I - \mu (H \otimes {I_q})}
\end{array}} \right] \cdot \tilde \Pi  - \tilde \Pi  \cdot {I_{qN}} + \left[ {\begin{array}{*{20}{c}}
\begin{array}{l}
0\\
0
\end{array}\\
{\mu (H \otimes {I_q})}
\end{array}} \right].
\end{equation*}
In this case, $\tilde \Pi (t) = \tilde \Pi (t_k^+)= \tilde \Pi $, then we have the requirement (RF) is fulled.

Secondly, we prove the the stability requirement (SF) part is satisfied.

Let $\widetilde{A}_{D}(t)={\rm blockdiag}(A_{D1}(t),\dots,A_{Dn}(t))$,
$\widetilde{B}_{D}(t) ={\rm blockdiag} (B_{D1}(t),\dots,B_{Dn}(t))$, $\widetilde{S}(t) =I_N\otimes {S_{Di}}(t)$.\\
Further more, let feedback $\widetilde{K}_{1}={\rm blockdiag}(K_{11},\dots,K_{1n})$, $\widetilde{K}_{2}={\rm blockdiag}(K_{21},\dots,K_{2n})$ and states  $x = [x_1^T(t), \ldots ,x_N^T(t)]$, $\eta  = [\eta _1^T(t), \ldots ,\eta _N^T(t)]$, and $ \tilde w(t_{k})=\mathbf{1}_N\otimes w(t_{k})$.
Then (\ref{sampledc}) imply
\begin{equation}\label{tsample}
\left[ {\begin{array}{*{20}{c}}
{x(t)}\\
{\xi (t)}\\
{\eta (t)}
\end{array}} \right] = \left[ {\begin{array}{*{20}{c}}
{{{\tilde A}_D}(t) + {{\tilde B}_D}(t){{\tilde K}_1}}&0&{{{\tilde B}_D}(t){{\tilde K}_2}}\\
{{{\tilde K}_1}}&0&{\tilde K}_2\\
0&0&{{{\tilde S}_{Di}}(t)}
\end{array}} \right]\left[ {\begin{array}{*{20}{c}}
{x(t_k^ + )}\\
{\xi (t_k^ + )}\\
{\eta (t_k^ + )}
\end{array}} \right] + \left[ {\begin{array}{*{20}{c}}
{{{\tilde P}_{Di}}}\\
0\\
0
\end{array}} \right]\tilde {w}(t_k ),\quad t \in ({t_k},{t_{k + 1}}],
\end{equation}
With (\ref{snetworkd}) and (\ref{tsample}) we have the following matrix, coordinating to $M\phi(h)\in \mathcal{D}_1$ in lemma 2, we note
\begin{equation}\label{M1}
{\mathbf{M}_2} = \left[ {\begin{array}{*{20}{c}}
{{{\tilde A}_D}(h) + {{\tilde B}_D}(h){{\tilde K}_1}}&0&{{{\tilde B}_D}(h){{\tilde K}_2} \cdot [I - \mu (H \otimes {I_q})]}\\
{{{\tilde K}_1}}&0&{{{\tilde K}_2} \cdot [I - \mu (H \otimes {I_q})]}\\
0&0&{{{\tilde S}_{Di}}(h) \cdot [I - \mu (H \otimes {I_q})]}
\end{array}} \right].
\end{equation}
According to the properties of partitioned matrices the eigenvalues of $\lambda(\mathbf{M}_2)$ of (\ref{M1}) are the eigenvalues of $\left[ {\begin{array}{*{20}{c}}
{{{\tilde A}_D}(h) + {{\tilde B}_D}(h){{\tilde K}_1}}&0\\
{{{\tilde K}_1}}&0
\end{array}} \right]$ and ${{{\tilde S}_{Di}}(h) \cdot [I_{qN} - \mu (H \otimes {I_q})]}$.\\
Which has been mentioned before, Assumption 3 implies $(A_{Di}(h), B_{Di}(h))$ are controllable.
Thus, for each agents $i,(i\in1,\dots,N)$, there exist  $K_{1i}$ such that ${{A_{Di}}(h) + {B_{Di}}(h){K_{1i}}}$ are Shur stable.
Then we have
\begin{equation}\label{eigup}
\sigma \left( {\left[ {\begin{array}{*{20}{c}}
{{{\tilde A}_D}(h) + {{\tilde B}_D}(h){{\tilde K}_1}}&0\\
{{{\tilde K}_1}}&0
\end{array}} \right]} \right) < 0.
\end{equation}

Since ${\tilde S_{Di}}(t) = \exp \left( {\tilde S \cdot h} \right)$, we have
\begin{equation*}
{{{\tilde S}_{Di}}(h) \cdot [I - \mu (H \otimes {I_q})]} = [I_N \otimes \exp(S\cdot h)]\cdot[I_{qN} - \mu (H \otimes {I_q})]=\exp (S \cdot h) \otimes({I_N} - \mu H)  .
\end{equation*}
With  the result in (\ref{21}), and (\ref{29}) is satisfied,  then we have
\begin{equation}\label{eigdown}
\sigma \left( {{{\tilde S}_{Di}}(h) \cdot [I - \mu (H \otimes {I_q})]} \right)<1.
\end{equation}
Then with (\ref{eigup}) and (\ref{eigdown}) are established, we have
\begin{equation*}
\sigma \left( {{\mathbf{M}_2}} \right) <1
\end{equation*}
Then we have the SF condition is satisfied for the system (\ref{snetworkc}) and (\ref{snetworkd}).

With the RF and SF condition satisfied for the system (\ref{snetworkc}) and (\ref{snetworkd}), according to Lemma 2, we can conclude that there exist an invariant set $\mathcal{M}=\{z|z(t)=\tilde \Pi(t) \tilde w(t)\}$ for
system (\ref{snetworkc}) and (\ref{snetworkd}), and it is is globally exponentially stable for system (\ref{snetworkc}) and (\ref{snetworkd}).

Moreover, with $0 = {C_{i}}{\Pi _{i}} + {Q_{i}}$, we have
\begin{equation}
 \mathop {lim}\limits_{t \to \infty } {e_i}(t)  = 0, \quad  i \in 1,2,...,N.
\end{equation}
The proof is completed.

\section{Simulation example}

In follows section, we provide two example to illustrate the cooperative output regulation with sampled compensator.
\subsection{Numerical simulation example}
Firstly, we consider cooperative output regulation for five agents with reference signal generated by an an unforced harmonic oscillator.
The five agents with dynamics given by (\ref{5}) and their parameters given by the following matrices
\begin{equation*}
{A_i} = \left[ {\begin{array}{*{20}{c}}
0&1&0\\
0&0&1\\
{ - 1}&2&3
\end{array}} \right]\quad {B_i} = \left[ {\begin{array}{*{20}{c}}
0\\
0\\
1
\end{array}} \right],\quad{\rm for}\quad i=1,\dots,5
\end{equation*}
and
\begin{equation*}
C = \left[ {\begin{array}{*{20}{c}}
1&1&1
\end{array}} \right],\quad Q = -C\Pi_i,\quad P = \left[ {\begin{array}{*{20}{c}}
0&0\\
0&0\\
0&1
\end{array}} \right],\quad {\rm for}\quad i = 1, \ldots ,5.
\end{equation*}
The reference system is described by the following equation
\begin{equation*}
\dot w(t) = \left[ {\begin{array}{*{20}{c}}
0&{ - 2}\\
2&0
\end{array}} \right]w(t).
\end{equation*}
The topology of the network is the same as the example in \cite{huangjie2012}, for your convenience, it is shown in fig. 1.
\begin{figure}[!htb]
\centering
\includegraphics[width=0.3\hsize]{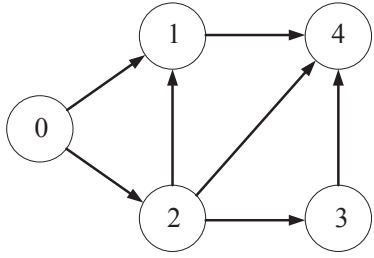}
\caption{\label{fig2} Observed value of $B'_{ij}$}
\end{figure}

The Laplacian matrix for the subgraph for the follower nodes (containing only node 1,2,3,4 and the connection between them) is given as
\begin{equation*}
\mathcal{L }= \left[ {\begin{array}{*{20}{c}}
1&{ - 1}&0&0\\
0&0&0&0\\
0&{ - 1}&1&0\\
{ - 1}&{ - 1}&{ - 1}&3
\end{array}} \right],
\end{equation*}
and the connection from node 0 is as follows
\begin{equation*}
\Delta  = \left[ {\begin{array}{*{20}{c}}
1&0&0&0\\
0&1&0&0\\
0&0&0&0\\
0&0&0&0
\end{array}} \right].
\end{equation*}

It is obviously that Assumptions from 1 and 4 are satisfied. There exist
\begin{equation*}
 {\Pi _i} = \left[ {\begin{array}{*{20}{c}}
{ - 0.0901}&{ - 0.0976}\\
{ - 0.1951}&{0.1801}\\
{0.3603}&{0.3903}
\end{array}} \right],
\end{equation*}
which satisfied ${\Pi_i }S = {A_i}{\Pi_i } +P_i$.\\
According to Theorem 2, the distributed sampled data compensator of the form (\ref{ithcloseloop2}) can solve the CORP. In (\ref{ithcloseloop2}), the feedback $K_{1i}$(for $i=1,\dots,4$) are express as
\begin{equation*}
  K_{1i}= [-8.9637,-10.3322,-10.7802;];
\end{equation*}
Additionally, we choose the sampling interval as $0.1s$, and $\mu = 0.1$ relevant to the sampling interval and topology of $\mathcal{G}$.

\begin{figure}
\begin{minipage}{0.40\linewidth}
  \centerline{\includegraphics[width=8.0cm]{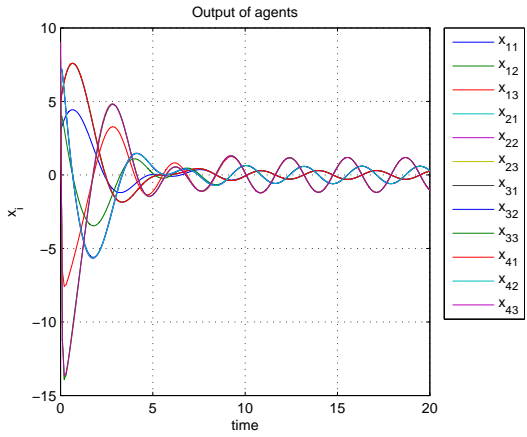}}
  \centerline{(a) Output of agents $x_i$}
\end{minipage}
\hfill
\begin{minipage}{0.40\linewidth}
  \centerline{\includegraphics[width=8.0cm]{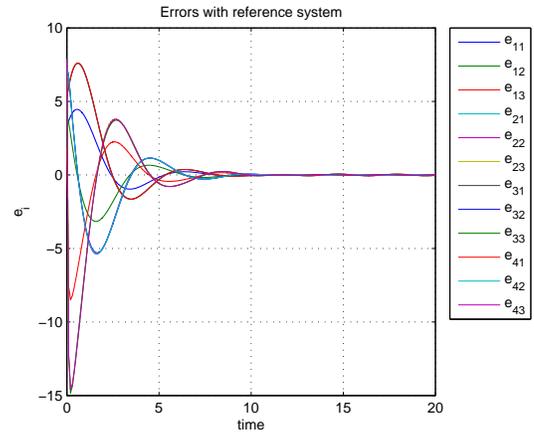}}
  \centerline{(b) Errors $e_i$}
\end{minipage}
\vfill
\begin{minipage}{0.40\linewidth}
  \centerline{\includegraphics[width=8.0cm]{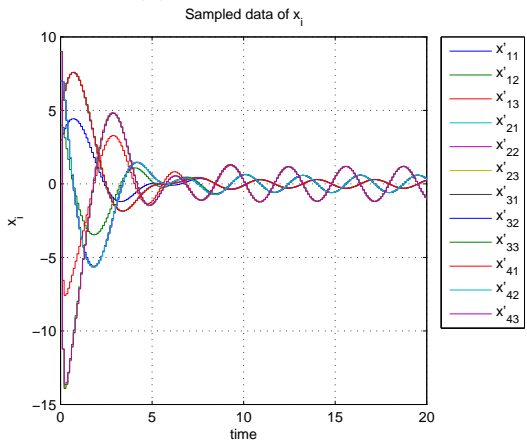}}
  \centerline{(c) Sampled data $x'_i$}
\end{minipage}
\hfill
\begin{minipage}{0.40\linewidth}
  \centerline{\includegraphics[width=8.0cm]{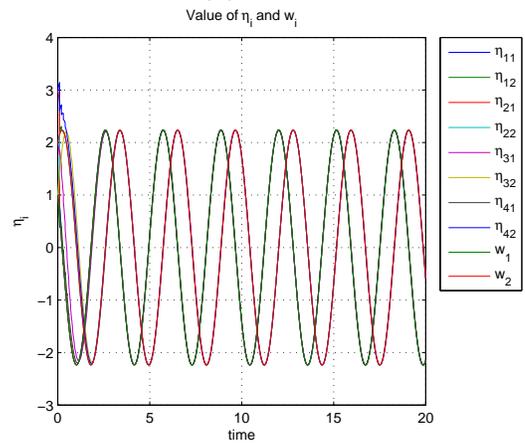}}
  \centerline{(d) Trajectory of $\eta_i$ and $w_i$}
\end{minipage}
\caption{Example results.}
\label{fig:result}
\end{figure}

Initial state $x_i(0)$ and $w_i(0)$ and $\eta_i(0)$ are random chosen.

The simulation results are represent by fig. 2, they represent the errors with the reference system and output of each system. Fig.2(a) shows the output of agents 1 to 4, it can be seen that their output $Cx_i$ converge to the same trajectory which generated by the reference signal $\Pi_i w$. Fig.2(b) shows that the tracking errors of all
the agents converge to zero. The sampled data of $x_i$ is shown in Fig.2(c) which is used in the feedback controller. The trajectory of dynamic compensators $\eta_i$ is shown in Fig.2(d), in Fig.2(d) it can be seen that they converge to the reference signal $w$.

\subsection{An application example in economic dispatch and frequency control of micro grid}
We introduce this example to illustrate why the sampled compensator for CORP is needed.

As mentioned in \cite{Mojica}, to improve the overall micro-grids (MGs) performance, i.e., reliability and effectiveness, a control scheme that combines the advantages of economic dispatch and frequency control is a hierarchical architecture. It can be express by the fig.3.
\begin{figure}[htbp]
\centering
  \includegraphics[width=0.8\textwidth]{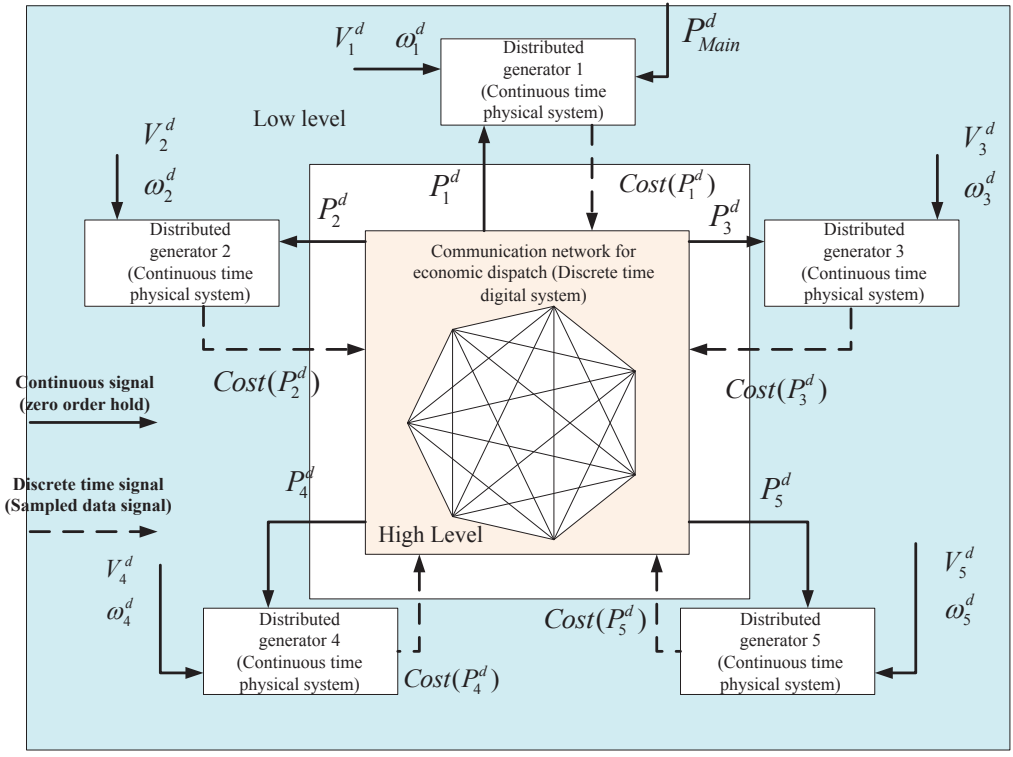}
  \caption{\label{fig3} line outage detection test in 39buses}
  \label{hierarchical architecture 1}
\end{figure}

At the high level of the MGs architecture, we apply a distributed strategy for the economic dispatch problem (EDP), which has been introduced detailedly in \cite{zhang}. The EDP belongs to the resource dispatch problem \cite{Wollenberg} which can be solved by the discrete incremental cost consensus algorithm. Here we rephrase the discrete algorithm and the example in section V-B from \cite{zhang} into the following equations
\begin{equation}\label{LD}
{\Lambda _i}(t_k^ + ) = {\Lambda _i}({t_k}) - \left( {{\mu _i} \cdot \sum\limits_{j \in {\mathcal{N}_i}} {{a_{ij}}({\Lambda _i}({t_k}) - {\Lambda _j}({t_k}))}  + {a_{i0}}(\sum\limits_{i = 1}^N {P_i^{r}} ({t_k}) - P_{main}^{r}({t_k}))} \right),{\rm for}\quad i,j = 1, \ldots ,5,
\end{equation}
and
\begin{equation}\label{PD}
P_i^{r}({t_k^+}) =\frac{1} {\alpha _i}({\Lambda _i}(t_k^+ ) - {\beta _i}),{\rm for}\quad i,j = 1, \ldots ,5,
\end{equation}
where $\Lambda _i(t_k^+)$ notes the incremental cost(IC) of $i$th MGs, $P_i^{r}$ denotes the energy desired by $i$th MGs.
$\alpha _i$  and $\beta _i$ are the coefficients of incremental cost function of $i$th MGs. For each MGs, the value of $\mu_i$ are taken as $\sum\limits_{j = 1}^n {\left| {{l_{ij}}} \right|},j=1,\dots,N$.
$P_{main}^{r}$ is the power desired by the main grid which can only be accessed by the micro-grid 1 in this example (i.e. $a_{10} = 0.0005$, otherwise $a_{i0} = 0$). The value of these parameters are list in table 1.

\begin{table}
 \caption{\label{table2} The value of parameters of outage lines}
 \centering
 \begin{tabular}[htb!]{ccccc}
  \toprule
  MGs  & $\alpha_i$ & $\beta_i$ &  $P^r_i(0)$  & $a_{i0}$\\
  \midrule
   1           &  561       &7.92       &200           &0.0005\\
   2           &  310       &7.85	    &150           &0\\

   3           &78          &7.8        &100           &0\\

   4           &561         &7.92	    &100           &0\\

   5           &78          &7.8	    &100           &0\\
  \bottomrule
 \end{tabular}
\end{table}

The Laplacian matrix $\mathcal{L}_c$ of the communication network is as follows
\begin{equation*}
 {\mathcal{L}_c} = \left[ {\begin{array}{*{20}{c}}
4&{ - 1}&{ - 1}&{ - 1}&{ - 1}\\
{ - 1}&1&0&0&0\\
{ - 1}&0&1&0&0\\
{ - 1}&0&0&1&0\\
{ - 1}&0&0&0&1
\end{array}} \right].
\end{equation*}

At the low level of the MGs architecture, the physical system of $i$th inverter-based micro grid ($i=1,2,...,n$) can be represented as the following model \cite{Schiffer}
\begin{equation}\label{modleS}
\begin{array}{*{20}{l}}
{{{\dot \delta }_i}(t) = {\omega _i}(t),}\\
{{{\dot \omega }_i}(t) =  - \frac{1}{{{\tau _{pi}}}}({\omega _i}(t) - \frac{1}{{{\tau _{pi}}}}{\omega ^d}) - \frac{{{k_{pi}}}}{{{\tau _{pi}}}}({P_i}(t) - P_i^d(t)),}\\
{{{\dot V}_i}(t) =  - \frac{1}{{{\tau _{Vi}}}}({V_i}(t) - \frac{1}{{{\tau _{Vi}}}}{V_i^d}) - \frac{{{k_{Vi}}}}{{{\tau _{pi}}}}({Q_i}(t) - Q_i^d(t)).}
\end{array}\quad t \in ({t_k},{t_{k + 1}}]
\end{equation}
the physical meaning of the variable can be found in \cite{Schiffer}. $\delta_i$ notes the phase angle of $i$th MG, $\omega_i$ notes the frequency of $i$th MG and $V_i$ notes the frequency of $i$th MG. $P_i$ is the active power of $i$th MG, $Q_i$ is the reactive power of $i$th MG. $\omega^d$ notes the desired frequency of all MGs which is the same frequency of main grid and $V_i^d$ notes the desired voltage of $i$th MG. They optimal value are both $\frac{{\sqrt 2 }}{2}P_i^r(t_k^+)$ \cite{Mojica}, where $P_i^r(t_k^+)$ are generated by (\ref{LD}) and (\ref{PD}). $\tau _{pi}$ and $\tau _{Vi}$ are the time constant of a low-pass filter.

More over we introduce differential control for $ P_i$ and $Q_i$ as follows
\begin{equation}\label{dc}
\left\{ \begin{array}{l}
{{\dot P}_i}(t) =  - {k_1}({P_i}(t) - P_i^d(t)) =  - {k_1}({P_i}(t) - \frac{{\sqrt 2 }}{2}P_i^d(t)),\\
{{\dot Q}_i}(t) =  - {k_2}({Q_i}(t) - Q_i^d(t)) =  - {k_1}({P_i}(t) - \frac{{\sqrt 2 }}{2}P_i^d(t)).
\end{array} \right.\quad t \in ({t_k},{t_{k + 1}}]
\end{equation}
Combine (\ref{modleS}) and (\ref{dc}) with zero-order hold for $P_i^r(t) = P_i^r(t_k^+)$, letting $\Delta {\omega _i}(t)=  {\omega _i}(t) - \omega ^d $, $\Delta {{V}_i}(t)={{ V}_i}(t) - {{V}_i^d}(t)$, then we have
\begin{equation}\label{cont}
\frac{d}{{dt}}\left[ {\begin{array}{*{20}{c}}
{ {\delta _i}(t)}\\
{\Delta {\omega _i}(t)}\\
{\Delta {{\dot V}_i}(t)}\\
{{P_i}(t)}\\
{{Q_i}(t)}
\end{array}} \right] = \left[ {\begin{array}{*{20}{c}}
0&1&0&0&0\\
0&{ - \frac{1}{{{\tau _{pi}}}}}&0&{ - \frac{{{k_{pi}}}}{{{\tau _{pi}}}}}&0\\
0&0&{ - \frac{1}{{{\tau _{pi}}}}}&0&{ - \frac{{{k_{qi}}}}{{{\tau _{pi}}}}}\\
0&0&0&{ - {k_1}}&0\\
0&0&0&0&{ - {k_2}}
\end{array}} \right]\left[ {\begin{array}{*{20}{c}}
{ {\delta _i}(t)}\\
{\Delta {\omega _i}(t)}\\
{\Delta {{\dot V}_i}(t)}\\
{{P_i}(t)}\\
{{Q_i}(t)}
\end{array}} \right] + \left[ {\begin{array}{*{20}{c}}
0\\
{\frac{{\sqrt 2 {k_{pi}}}}{{2{\tau _{pi}}}}}\\
{\frac{{\sqrt 2 {k_{qi}}}}{{2{\tau _{pi}}}}}\\
{ - \frac{{\sqrt 2 }}{2}{k_1}}\\
{ - \frac{{\sqrt 2 }}{2}{k_2}}
\end{array}} \right]P_i^r(t_k^ + ),\quad t \in ({t_k},{t_{k + 1}}]
\end{equation}

It is easy to obtain that SF condition is full filled for the above system (\ref{LD}) and (\ref{cont}) with positive $\tau _{pi}$, $\tau _{vi}$, $k_1$, $k_{pi}$ and $k_{qi}$. According to (\ref{LD}), $S=0$ for the reference system, then there always exist $\Pi_i(t)=0$, and an unique $\Pi_i(t_k^+)$ satisfied make the RF condition is full filled. With theorem 1, the CORP of (\ref{LD}) and (\ref{cont}) can be achieved. Moreover, in this case, the problem is degenerated as cooperative tracking for an constant value.
\begin{figure}
\begin{minipage}{0.40\linewidth}
  \centerline{\includegraphics[width=8.0cm]{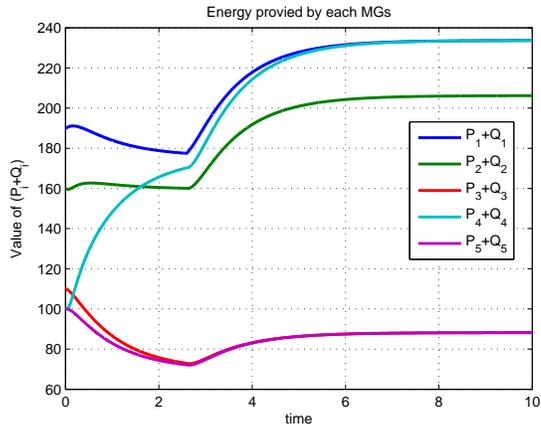}}
  \centerline{(a) $P_i+Q_i$}
\end{minipage}
\hfill
\begin{minipage}{0.40\linewidth}
  \centerline{\includegraphics[width=8.0cm]{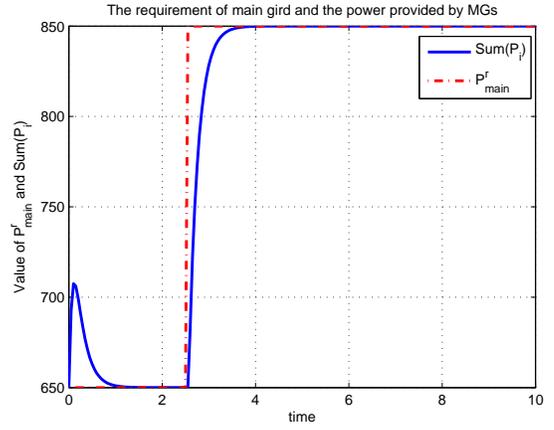}}
  \centerline{(b) $P_{main}^d$ and $\sum\limits_{i = 1}^n {P_i+Q_i} $}
\end{minipage}
\vfill
\begin{minipage}{0.40\linewidth}
  \centerline{\includegraphics[width=8.0cm]{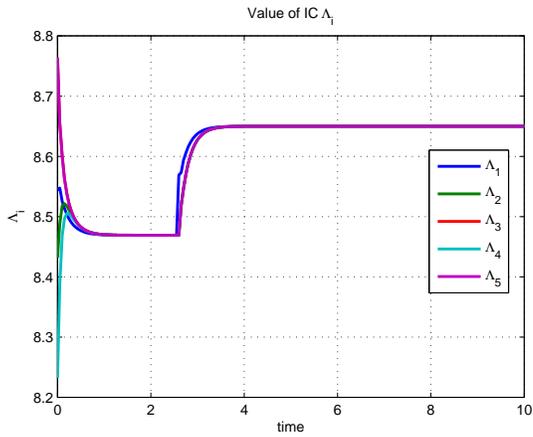}}
  \centerline{(c) IC $\Lambda_i$}
\end{minipage}
\hfill
\begin{minipage}{0.40\linewidth}
  \centerline{\includegraphics[width=8.0cm]{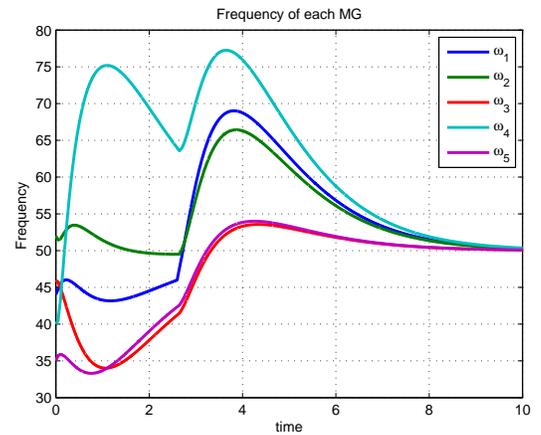}}
  \centerline{(d) Frequency $\omega_i$ of MGs}
\end{minipage}
\caption{Example results}
\label{fig:res}
\end{figure}

The economic dispatch and frequency control result are shown in fig.4. The units of time is second (s), the units of energy $R^r_{main}$ and $P^d_i+Q^d_i$ are MWh, the units of incremental cost is \$/MWh , the units of frequency is Hz.

Fig.4(a) shows the energy provided by each MG. Fig.4(b) shows the summation of the energy provided by MGs even the energy requirement of main grid, the requirement with an initial value of 650 and change to 850 at 2.3s. Fig.4(c) shows the IC converge to a common value which means the economic dispatch problem is solved. Finally, Fig.4(d) shows the frequency of each MG is converging to the frequency of main grid.

\section{Conclusion}
In this work, we have investigated the cooperative output regulation problem
of continuous-time multi-agent systems with sampled data compensator and discrete-time communication. Firstly, we have discussed the problem generally via a hybrid internal model principle method. Then we presented a detailed sampled data compensator designing method for this problem.

In this paper, the ripple problem is not taken into consideration, so there are still some ripples in the trajectory of errors in fig.2(b). It will be our first consideration in future work.
Another future work need to be done is that we should extend the research to systems with parameter uncertainty and extra disturbance not generated by the reference system.
Moreover, research of controller with only output information is needed to be done, too.

In general, more effort should be given in the research of CPS. The CPS consists of physical plants cooperative with each others with digital communication will be very common whose signal flow has the hybridity in time. And the research of CPS modeled by hybrid system will be more and more popular in future.

\end{document}